\documentclass[aip,reprint,longbibliography,superscriptaddress]{revtex4-2}
\usepackage{graphicx,color}
\usepackage{amsmath, amssymb}
\usepackage{longtable}
\usepackage{natbib}
\usepackage{bm}% bold math
\usepackage[hidelinks]{hyperref}% add hypertext capabilities
\hypersetup{
  colorlinks   = true, %Colours links instead of ugly boxes
  urlcolor     = blue, %Colour for external hyperlinks
  linkcolor    = blue, %Colour of internal links
  citecolor   = red %Colour of citations
}
\usepackage[utf8]{inputenc}

\begin{document}
\title{Breakdown of broken-symmetry approach to exchange interaction}
\author{Naoya Iwahara}
\email{naoya.iwahara@gmail.com}
\affiliation{Graduate School of Engineering, Chiba University, 1-33 Yayoi-cho, Inage-ku, Chiba-shi, Chiba 263-8522, Japan} 
\affiliation{Theory of Nanomaterials Group, KU Leuven, Celestijnenlaan 200F, B-3001 Leuven, Belgium}
\author{Zhishuo Huang}
\affiliation{Theory of Nanomaterials Group, KU Leuven, Celestijnenlaan 200F, B-3001 Leuven, Belgium}
\affiliation{Department of Chemistry, National University of Singapore, Block S8 Level 3, 3 Science Drive 3, 117543, Singapore}
\author{Akseli Mansikkam\"{a}ki}
\affiliation{NMR Research Unit, University of Oulu, P.O. Box 3000, FI-90014 Oulu, Finland}
\author{Liviu F. Chibotaru}
\email{liviu.chibotaru@kuleuven.be}
\affiliation{Theory of Nanomaterials Group, KU Leuven, Celestijnenlaan 200F, B-3001 Leuven, Belgium}
\date{\today}

\begin{abstract}
Broken-symmetry (BS) approaches are widely employed to evaluate Heisenberg exchange parameters, primarily in combination with DFT calculations. 
For many magnetic materials, BS-DFT calculations give reasonable estimations of exchange parameters, although systematic failures have also been reported.
While the latter were attributed to deficiencies of approximate exchange-correlation functional, we  
prove here by treating a simple model system that the broken-symmetry methodology has serious problems. 
Detailed analysis clarifies the intrinsic issue with the broken-symmetry treatment of low-spin states.
It shows, in particular, that the error in the BS calculation of exchange parameter scales with the degree of covalency between the magnetic and the bridging orbitals. This is due to the constraint on the form of multiconfigurational state imposed by the BS determinant, a feature common to other single-reference methods too.
As a possible tool to overcome this intrinsic drawback of single-determinant BS approaches, we propose their extension to a minimal multiconfigurational version.
\end{abstract}

\maketitle

\section{Introduction}
In the study of magnetism, understanding the exchange mechanisms and evaluating the exchange coupling parameters ($J$) are of fundamental importance \cite{Anderson1963, Goodenough1963, Ginsberg1971, Geertsma1990, Kahn1993, Khomskii2014, Streltsov2017}.
Current quantum chemistry methodologies offer approaches for evaluating the exchange parameters in various magnetic molecules and materials of experimental interest \cite{Ceulemans2000, Ruiz2004, Neese2009, Xiang2013, Riedl2019}.
Among them, the broken-symmetry (BS) approach \cite{Noodleman1981, Yamaguchi1986, Soda2000} has become a standard tool, especially for calculating $J$ in large 
systems involving several magnetic centers. This method allows the extraction of the exchange coupling by relating the energies of high-spin and BS low-spin states to the corresponding spin configurations of the Heisenberg exchange model. This approach is universal and can be applied with 
various quantum chemistry methods allowing for calculating the total energy of different electronic configurations. Most calculations of this type 
have been performed within density functional theory (DFT) and, to a much lesser extent, within Hartree-Fock approximation. Recently, calculations of exchange parameters with BS-coupled cluster (CC) \cite{BS_CC}, BS-$G_0W_0$ \cite{BS_GW}, and BS-self-consistent GW \cite{Pokhilko2022, Pokhilko2023} method have also been reported. 

Despite its simplicity, the BS-DFT approach provides reasonable exchange parameters, often close to experimental values, for many magnetic materials \cite{Martin1997, Ruiz2004, Neese2009, Riedl2019}.
Simultaneously, many quantitative and qualitative failures have been reported and discussed \cite{Ruiz1997, Akseli_PhD}.
In particular, the strong dependence of the calculated $J$ on the exchange-correlation functionals in the DFT calculation was found \cite{Rivero2008, Valero2008, Peralta2010, Phillips2011, Phillips2012, Akseli_PhD}. 
Although the failures have often been attributed to the limitations of quantum chemistry methods \cite{Polo2003}, these may not be the sole reason.

Here, we prove the breakdown of the broken-symmetry approach for extracting $J$ by considering a generic three-site model. 
We demonstrate that this breakdown originates from an artificial constraint on the multiconfigurational state imposed by the broken-symmetry determinant. 
The error becomes especially pronounced in the case of strong covalency between the magnetic centers and the bridging ligands.
To overcome this drawback, we propose a calculational scheme based on a minimal version of the multiconfigurational (MC) BS approach. 

\section{Generic three-site model}
\label{Sec:model}
To assess the performance of the BS method, we confront its prediction for the exchange parameter $J$ with the results of exact diagonalization. 
To facilitate the subsequent analysis, we treat the simplest possible model that contains all necessary physical ingredients. 

\begin{figure}
\begin{tabular}{lll}
(a) & ~~ & (b) \\
\includegraphics[width=0.45\linewidth]{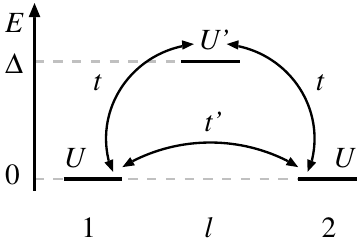}
&& 
\includegraphics[width=0.45\linewidth]{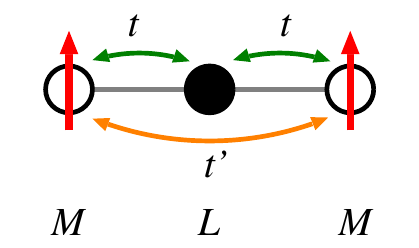}
\\
(c) && (d) \\
\includegraphics[width=0.45\linewidth]{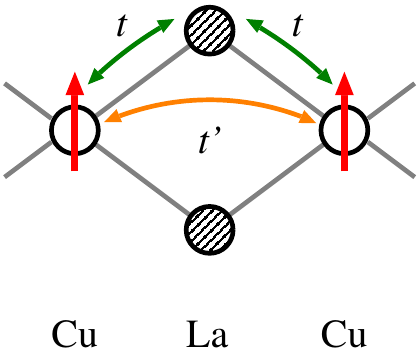}
&&
\includegraphics[width=0.45\linewidth]{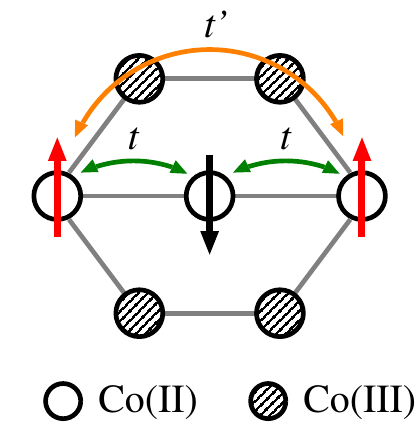} 
\end{tabular}
\caption{Three-site systems.
(a) Parametrization. 
The geometries of 
(b) trinuclear complex such as [LFeCoFeL]$^{3+}$ \cite{Glaser1999}, 
(c) Cu chain \cite{Mizuno1990, Masuda1991},
and 
(d) Co$^\text{II}_3$Co$^\text{III}_4$ complex \cite{Chibotaru2008}. 
}
\label{Fig:system}
\end{figure}

We consider a three-center system consisting of two half-filled magnetic orbitals (1,2) and one empty orbital ($l$) at the bridging ligand group [see Fig. \ref{Fig:system}(a)]. 
The two particles are either electrons or holes, depending on the situation. 
While looking just as H-He-H on a minimal basis, such a model can describe real magnetic materials if proper orbitals 1, 2, and $l$ are chosen \cite{3center}.

The model possesses inversion/reflection symmetry with respect to the ligand site and includes the nearest-neighbor electron transfer ($t$) between the magnetic and the ligand orbital and the next-nearest neighbor electron transfer ($t'$) between orbitals 1 and 2.
Adding electron repulsion on sites, we end up with a $t-t'$ Hubbard model: 
\begin{eqnarray}
 \hat{H} &=& \sum_{\sigma = \uparrow, \downarrow} t 
 (\hat{a}_{1\sigma}^\dagger \hat{a}_{l\sigma} + \hat{a}_{l\sigma}^\dagger \hat{a}_{1\sigma} 
 + \hat{a}_{2\sigma}^\dagger \hat{a}_{l\sigma} + \hat{a}_{l\sigma}^\dagger \hat{a}_{2\sigma})
\nonumber\\
 &&+ \sum_{\sigma = \uparrow, \downarrow} t' 
 (\hat{a}_{1\sigma}^\dagger \hat{a}_{2\sigma} + \hat{a}_{2\sigma}^\dagger \hat{a}_{1\sigma})
 + \sum_{\sigma = \uparrow, \downarrow} \Delta \hat{n}_{l\sigma}
\nonumber\\
 &&+ U(\hat{n}_{1\uparrow} \hat{n}_{1\downarrow} + \hat{n}_{2\uparrow} \hat{n}_{2\downarrow})
   + U' \hat{n}_{l\uparrow} \hat{n}_{l\downarrow}.
\label{Eq:H}
\end{eqnarray}
Here, $i$ indicates orbitals, $\sigma$ is the electron spin projection, $\hat{a}_{i\sigma}^\dagger$ and $\hat{a}_{i\sigma}$ are 
electron/hole creation and annihilation operators in the spin-orbital $i\sigma$, respectively; $\hat{n}_{i\sigma} = \hat{a}_{i\sigma}^\dagger 
\hat{a}_{i\sigma}$,
$\Delta$ is the gap between metal and ligand orbital levels, and $U$ and $U'$ are the parameters of Coulomb repulsion in the magnetic and ligand orbitals, respectively. 
We assume the orthonormality of involved orbitals, $\langle i|j\rangle = \delta_{ij}$.
The sign of $t$ does not influence the energy levels; hence, hereafter, $t \ge 0$.

Despite the simplicity, this model reproduces all essential contributions to the exchange interaction between two unpaired spins except for the spin polarization of ligands. 
The inclusion of direct electron transfer ($t'$) along with the intermediate one ($t$) is indispensable for quantitative analysis of exchange interaction in many magnetic materials \cite{3center}. 
Besides conventional kinetic antiferromagnetic and potential ferromagnetic contributions \cite{Anderson1959, Anderson1963}, it also allows us to identify the ferromagnetic kinetic exchange contribution \cite{Tasaki1995, Chibotaru1996, Chibotaru2003, Penc1996, Tasaki2020}. 
The latter plays a vital role in complexes with strong metal-ligand covalency, such as the thiophenolate-bridged heterotrinuclear complex [LFeCoFeL]$^{3+}$ \cite{Glaser1999}, in which one orbital of the bridging ligand group containing the diamagnetic Co(III) is strongly hybridized with the magnetic orbitals at low-spin Fe(III) sites \cite{Chibotaru1996, Chibotaru2003, 3center}. 
Such a generic situation [Fig. \ref{Fig:system}(b)] is met for numerous bridging groups L, the limitation to one empty or doubly occupied ligand orbital $l$ being sufficient in many cases \cite{3center}.
For the bridging geometry in Fig. \ref{Fig:system}(b), $t'$ is expected to be significantly smaller than $t$. 
However, for strong metal-ligand covalency, the former can be far from negligible as in iron-sulfur bridged [LFeCoFeL]$^{3+}$ where $t'=0.4|t|$ \cite{3center}. 
At the same time, we can easily conceive M-L-M structures where $t'$ can be of similar magnitude with or even larger than $t$ [Fig. \ref{Fig:system}(c),(d)]
\footnote{
More precisely, $\unexpanded{t'}$ should be compared with $\unexpanded{t^2/\Delta}$ ($\unexpanded{\Delta \gg |t|, |t'|}$) \cite{3center}. 
When they are comparable, their effect on the exchange interaction is similar.
}.

Given the above arguments, a comparison of exact and BS calculations of $J$ based on this model is expected to be conclusive in the quest for the validity of the latter. 
Moreover, the identification of the domain of parameters of Eq. (\ref{Eq:H}) for which the discrepancy occurs will provide direct insight into the physical reasons for its breakdown 
\footnote{Although the model (\ref{Eq:H}) provides a potential exchange contribution when one passes from the Wannier orbitals 1, 2 and $l$ to the magnetic orbitals of Anderson's type \cite{vandenHeuvel2007}, appreciable direct potential exchange interaction can already exist for the localized orbitals 1 and 2 in many magnetic materials \cite{3center}. For simplicity, we do not include this interaction in the model (\ref{Eq:H}) since it does not affect the main conclusions of this work.
}.

\begin{figure*}
\begin{tabular}{lll}
(a) &~~& (b) \\
\includegraphics[width=0.48\linewidth]{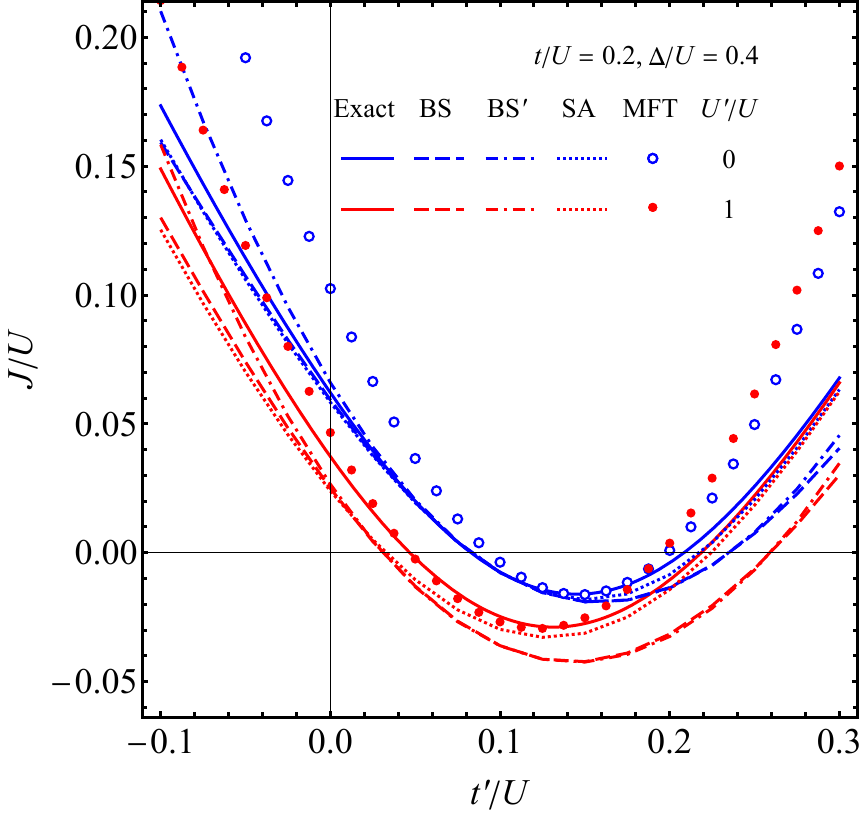}
&&
\includegraphics[width=0.48\linewidth]{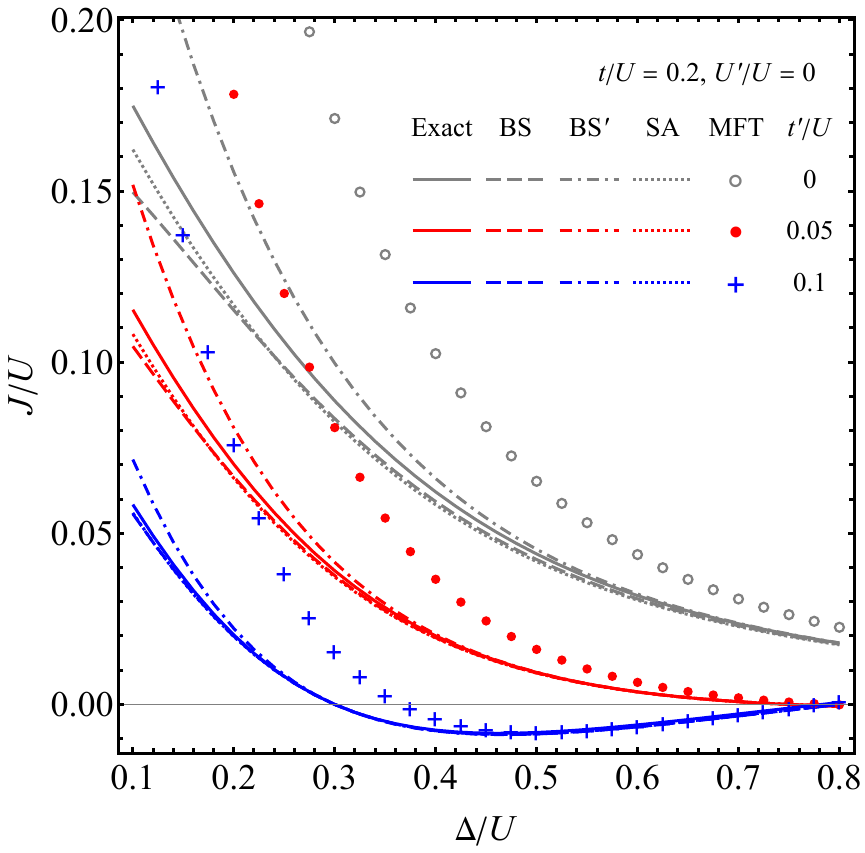}
\\
(c) & & (d) \\
\includegraphics[width=0.48\linewidth]{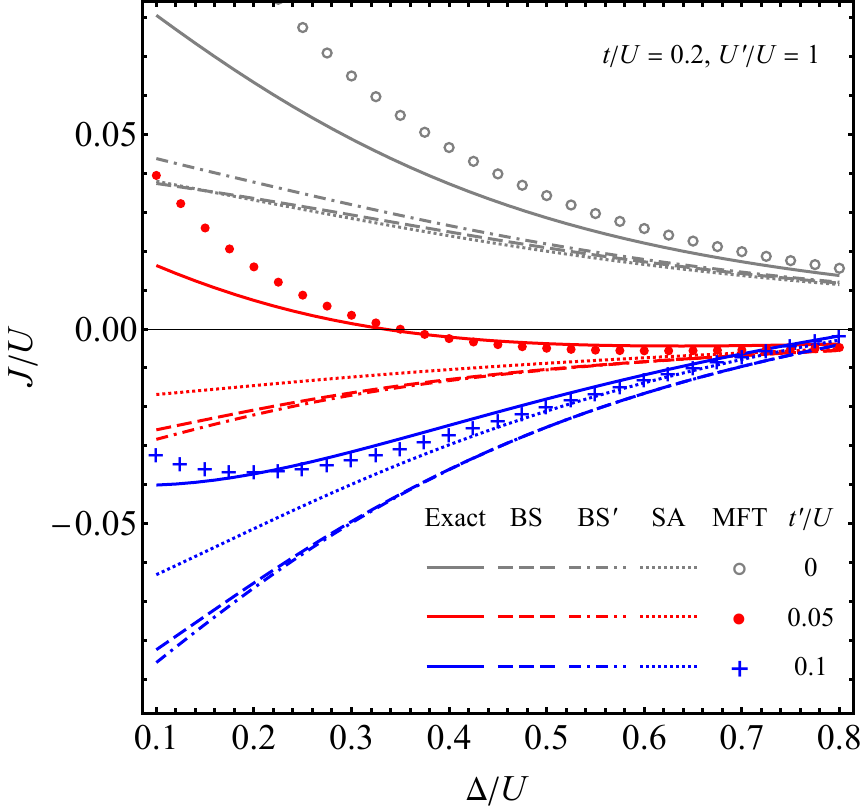}
&&
\includegraphics[width=0.48\linewidth]{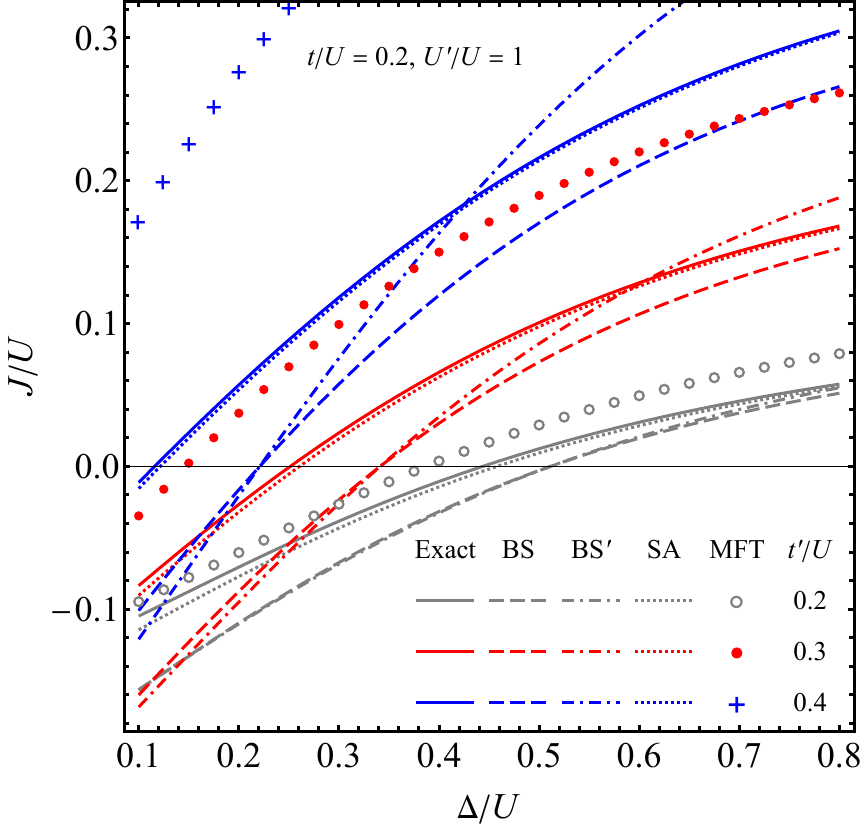}
\end{tabular}
\caption{
Exact and approximate exchange parameters. 
The exact $J$ (solid), $J_\text{BS}$ (dashed), $J_{\text{BS}'}$ (dot-dashed), $J_\text{SA}$ (dotted), and $J_\text{MFT}$ (symbols) with respect to (a) $t'/U$ and (b), (c), (d) $\Delta/U$. 
The parameters used for the simulations are written in the figure. 
(a) The blue and red indicate, respectively, $U'/U = 0$ and 1. 
(b)-(d) The gray, red, and blue indicate the data with different $t'/U$. Abbreviation used: BS - Yamaguchi's Eq. (\ref{Eq:Jbs}), BS' - Noodleman's Eq. (\ref{Eq:Jbs2}), SA - spin symmetry adapted expression (\ref{Eq:JSA}), MFT - magnetic force theorem (\ref{Eq:JMFT}).
}
\label{Fig:J}
\end{figure*}

\section{Exact vs broken-symmetry calculations of exchange parameters}

We derive the Heisenberg model,
\begin{eqnarray}
 \hat{H}_\text{ex} &=& J \hat{\bm{S}}_1 \cdot \hat{\bm{S}}_2,
\label{Eq:Hex}
\end{eqnarray}
from the ground ferromagnetic (high-spin) and antiferromagnetic (low-spin) energies of our three-center Hubbard model (\ref{Eq:H}).
In this equation, $\hat{\bm{S}}_i$ is a spin 1/2 operator on site $i=1,2$. 
We derive the exchange parameter $J$ using the full configuration interaction (CI) method (exact solution), and by using the broken-symmetry Hartree-Fock (BS-HF) approximation. 

\subsection{Exact diagonalization}
In the first approach, we determine the exchange parameter $J$ in the Heisenberg model using the energy gap between the exact ferromagnetic (high-spin) and antiferromagnetic (low-spin) energy levels obtained from the full CI calculations:
\begin{align}
 J = E_\text{F} - E_\text{AF}.
 \label{Eq:Jexact}
\end{align}
See for the details of the full CI Hamiltonian matrices Appendix \ref{A:exact}.

Figure \ref{Fig:J} shows the calculated exact $J$ (\ref{Eq:Jexact}) with solid lines. 
As previously demonstrated\cite{3center}, the exchange parameter can be ferromagnetic ($J<0$) and antiferromagnetic ($J>0$) depending on the microscopic interaction parameters in Eq. (\ref{Eq:H}). 
More information on the dependence of $J$ on the parameters of the microscopic model (\ref{Eq:H}) can be found in Ref. \onlinecite{3center}.

\subsection{Broken-symmetry approaches}

\subsubsection{Yamaguchi's approach}
To estimate the exchange parameter in Eq. (\ref{Eq:Hex}) with the BS-HF method, we use Yamaguchi's formula \cite{Yamaguchi1986}:
\begin{eqnarray}
 J_\text{BS} &=& \frac{2(E_\text{F} - E_\text{BS})}
 {\langle \hat{\bm{S}}^2 \rangle_\text{F} - \langle \hat{\bm{S}}^2 \rangle_\text{BS}},
\label{Eq:Jbs}
\end{eqnarray}
where $\hat{\bm{S}} = \hat{\bm{S}}_1 + \hat{\bm{S}}_2$, 
$\langle \hat{\bm{S}}^2 \rangle_\text{F} = 2$, $\langle \hat{\bm{S}}^2 \rangle_\text{BS}$ is the expectation value of $\hat{\bm{S}}^2$ for BS-HF (unrestricted HF) wave function $|\Psi_\text{BS}\rangle$, 
and $E_\text{F}$ and $E_\text{BS}$ are the ferromagnetic (high-spin) and broken-symmetry energies, respectively. 
See for the detailed expressions for the BS-HF calculations Appendix \ref{A:UHF}.

Figure \ref{Fig:J}(a) shows the calculated exchange parameters in  function of $t'/U$.
The solid and the dashed lines are the exact $J$ obtained from the full CI calculations and the broken-symmetry $J_\text{BS}$, respectively. 
They exhibit similar behavior, whereas $J_\text{BS}$ are quantitatively and, under certain ranges of parameters, qualitatively different from the exact one.  
$J_\text{BS}$ tends to overestimate the ferromagnetic contribution due to the contamination of the ferromagnetic component of about 40-50 \% in the BS-HF energy (see Appendix \ref{A:UHF}). 
The spin contamination in the BS-HF wave function makes the variational parameters (molecular orbital coefficients) not fully optimal for the description of the ground antiferromagnetic state.

Moreover, the number of variational parameters for the BS-HF wave function is smaller than for the exact solution, leading to a poorer  description of the antiferromagnetic state with the BS-HF wave function. 
For some range of $t'/U$, one can also see that $J_\text{BS}$ qualitatively differs (has opposite sign) from the exact $J$. Thus for $0.03 \alt t'/U \alt 0.05$ and $0.22 \alt t'/U \alt 0.26$ for $U'/U = 1$ ($0.21 \alt t'/U \alt 0.24$ for $U'/U = 0$), the exact $J$ becomes antiferromagnetic, whereas $J_\text{BS}$ is ferromagnetic 
\footnote{
The qualitative difference between $J$ and $J_\text{BS}$ appears due to the explicit treatment of the ligand atom. 
Within the two-site Hubbard model, the behavior of $J$ and $J_\text{BS}$ as function of the parameters is similar \cite{analytical}.
}.
The discrepancy is enlarged with the increase of the Coulomb repulsion on the bridging site, $U'$, implying 
that the mean-field description is not adequate. This means that the
static electron correlation involving explicitly ligand type configurations is crucial to derive accurate $J$.

We can see the importance of the electron correlation for the description of $J$ by modulating the metal-ligand covalency via $\Delta$. 
From a detailed analysis of the present model, we have demonstrated earlier that the static electron correlation effect is enhanced for small $\Delta$ \cite{3center}. 
Figures \ref{Fig:J}(b), (c) show that $J_\text{BS}$ deviates from exact $J$ when $\Delta$ is diminished and the covalency effects are enhanced. The discrepancy is further enhanced by turning on $U'$.

\subsubsection{Noodleman's or mapping approach}
\begin{figure}
\includegraphics[width=0.9\linewidth]{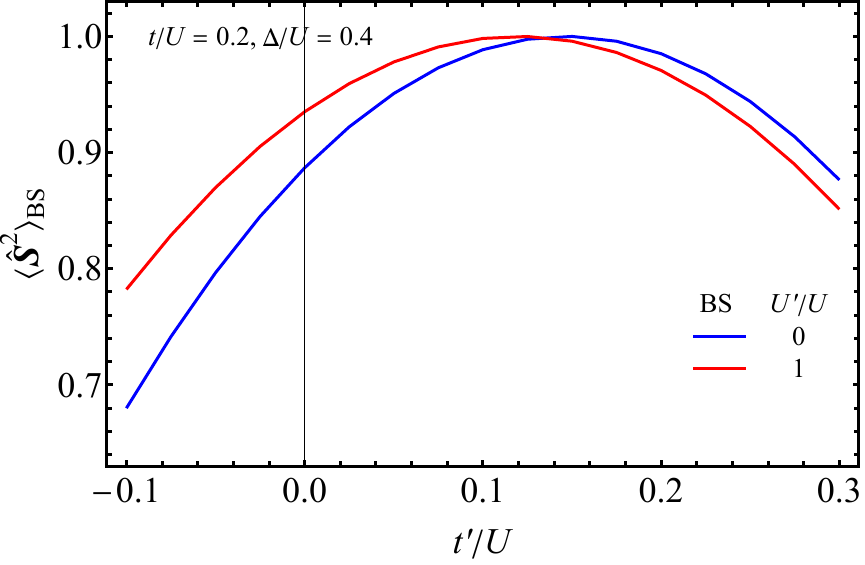}
\caption{
\label{Fig:S2}
$\langle \hat{\bm{S}}^2\rangle_\text{BS}$ with respect to $t'/U$.
The interaction parameters are the same as those for Fig. \ref{Fig:J}(a).
}
\end{figure}

When the BS state displays well-localized spin densities on sites, one can equally well use the Noodleman's expression for the exchange parameter (further denoted $J_{\text{BS}'}$) \cite{Noodleman1981}, 
\begin{align}
 J_{\text{BS}'} &= 2(E_\text{F} - E_\text{BS}).
\label{Eq:Jbs2}
\end{align}
It corresponds to orthogonal magnetic orbitals, implying $\langle \hat{\bm{S}}^2 \rangle_\text{BS} \approx 1$ in the Yamaguchi's expression (\ref{Eq:Jbs}). 
However, $\langle \hat{\bm{S}}^2 \rangle_\text{BS}$ gradually deviates from unity with the change of $t'$ and $\Delta$ [see Fig. \ref{Fig:S2}]. 
Accordingly, the prediction based on Noodleman's formula deviates from the result given by Eq. (\ref{Eq:Jbs}). 
In the limit of large M-L covalency, when the HF instability does not occur (the BS-HF determinant coincides with a restricted HF solution) so that $\langle \hat{\bm{S}}^2 \rangle_\text{BS} = 0$, the Noodleman's expression will be strongly in error while Eq. (\ref{Eq:Jbs}) still correct
\footnote{
In the {\it ab initio} calculations of complexes, contrary to the HF method, the DFT calculations partly include the electron correlation through the exchange-correlation functional, an effect more pronounced for pure than hybrid functionals \cite{Cremer2002}. 
This effect is accounted for by the denominator of Yamaguchi's expression which is simply proved by the fact that in the limit of exact exchange-correlation functional, when $\langle\hat{\bm{S}}^2 \rangle_\text{BS-DFT} = 0$, the corresponding Eq. (\ref{Eq:Jbs}) will correctly describe the energy difference between two states with definite spin.   
}.

One should mention that similar ideas are contained in the so-called mapping approach \cite{Illas2000, Moreira2006}. If one denotes
the high-spin and broken-symmetry spin states $|S_{1z}, S_{2z}\rangle$ as $|\uparrow, \uparrow\rangle$ and $|\uparrow, \downarrow\rangle$, respectively, then both approaches give for the diagonal matrix elements of the Heisenberg model 
\begin{align}
 \langle \uparrow, \uparrow| \hat{H}_\text{ex} |\uparrow, \uparrow\rangle &= J S_1 S_2, 
 \nonumber\\
 \langle \uparrow, \downarrow| \hat{H}_\text{ex} |\uparrow, \downarrow\rangle &= -J S_1 S_2, 
\end{align}
so that their difference $\Delta E_\text{map}$ corresponds to $2J S_1 S_2$.
In the case of $S_1=S_2=1/2$ this gives for $J$ Eq. (\ref{Eq:Jbs2}).

Figure \ref{Fig:J} shows the calculated broken-symmetry $J_{\text{BS}'}$ (the dot-dashed lines) within Noodleman's formula (\ref{Eq:Jbs2}). 
As the delocalization becomes stronger by reducing $\Delta$  [Fig. \ref{Fig:J}(b)] or by increasing $t'$ [Fig. \ref{Fig:J}(d)], the discrepancy between the $J_{\text{BS}'}$ and exact $J$ becomes larger. 
Overall, Yamaguchi's $J_\text{BS}$ tends to be closer to the exact $J$ than Noodleman's $J_{\text{BS}'}$.

\section{Alternative single-reference methods for calculating $J$}
At this point it is legitimate to inquire whether other single-reference based methods allow to overcome the drawbacks of the BS approach. Below we consider three of them, the spin symmetry adapted approach (SA), the method employing the magnetic force theorem (MFT) and the spin-flip approach (SF). 

\subsection{Spin symmetry adapted approach}
This method replaces the BS-HF state with a low-spin wave function which preserves the spin symmetry. 
Projecting out the AF ($S=0$) part from the BS-HF wave function, further identified as the SA-HF wave function, $|\Psi_\text{SA}\rangle$ (\ref{Eq:WF_SA}), we can calculate the exchange parameter in the same way as in the BS approach. Thus
we evaluate the exchange parameter $J_\text{SA}$ by replacing $E_\text{BS}$ and $\langle \hat{\bm{S}}^2 \rangle_\text{BS}$ in Eq. (\ref{Eq:Jbs}) with the expectation values for $|\Psi_\text{SA}\rangle$: 
\begin{align}
 J_\text{SA} 
 = \frac{2(E_\text{F} - E_\text{SA})}{\langle \hat{\bm{S}}^2 \rangle_\text{F} - \langle \hat{\bm{S}}^2 \rangle_\text{SA}}
 = E_\text{F} - E_\text{SA},
 \label{Eq:JSA}
\end{align}
where $E_\text{SA}$ is the low-spin state energy.
For our model (\ref{Eq:H}), we have $\langle \hat{\bm{S}}^2 \rangle_\text{F} = 2$ and $\langle \hat{\bm{S}}^2 \rangle_\text{SA} = 0$.
For detailed expressions of the wave functions and energy, see Appendix \ref{A:SA}.

According to our calculations (the dotted lines in Fig. \ref{Fig:J}), the spin-adapted approach partly cures the discrepancy between the exact $J$ and $J_\text{BS}$
\footnote{
In practice, similar attempts have been made using DFT, while clear improvement of the calculated $J$ for real molecules has not been seen \cite{illas_2007a}. 
}.
The agreement between $J$ and $J_\text{SA}$ is better for weak electron correlation on the ligand atom, i.e., for smaller $U'/U$ [Fig. \ref{Fig:J}(a)], and larger $t'/U$ [Fig. \ref{Fig:J}(a)] and $\Delta/U$ [Fig. \ref{Fig:J}(b)-(d)].
Although the SA-HF approach resolves the spin contamination problem in the BS approach, the SA wave function possesses smaller degrees of freedom compared with the exact wave function, which leads to the discrepancy between the exact $J$ and $J_\text{SA}$ when the electron correlation effects becomes stronger.

\begin{figure}[tb]
\includegraphics[width=0.4\linewidth]{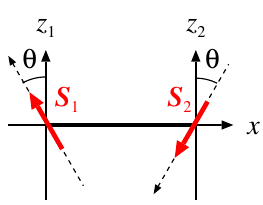}
\caption{Rotation of the spins for the magnetic force theorem calculations.}
\label{Fig:spin_cl}
\end{figure}

\subsection{Magnetic force theorem}
Another popular method for the calculation of exchange parameters is based on the use of magnetic force theorem \cite{Oguchi1983, Liechtenstein1984, Lichtenstein1987}.
In this approach, the ground-state BS energy, minimized under constrained directions of magnetization on magnetic sites, is confronted with a classical spin model with similar directions of the corresponding spins. 
Under the rotations of the antiparallel classical spins from initial collinear arrangement (Fig. \ref{Fig:spin_cl}), the total energy reads 
\begin{align}
 E(\theta) = -J_\text{MFT} \bm{S}^2 \cos 2\theta \approx -J_\text{MFT}  \bm{S}^2 \left(1 - 2 \theta^2\right),
 \label{Eq:Ecl_theta}
\end{align}
where $\bm{S}$ is the classical spin vector, and the last equality assumes a small $\theta$. 
Within the quantum mechanical treatment, the rotation of the quantum spins of the BS-HF state is achieved by $\hat{R}(\theta) |\Psi_\text{BS}\rangle$ with 
\begin{align}
 \hat{R}(\theta) = e^{i\hat{S}_{1y} \theta} e^{-i\hat{S}_{2y}\theta}.
 \label{Eq:Rtheta}
\end{align}
By comparing the classical and quantum mechanical energies, we obtain $J_\text{MFT} $. 
The expressions for the energy and $J_\text{MFT} $ are given in Appendix \ref{A:MFT}.

Fig. \ref{Fig:J}  (symbols) shows the resulting exchange parameter as function of $t'/U$ and $\Delta/U$. One can see that
$J_\text{MFT}$ shows opposite tendencies to $J_\text{BS}$ and $J_\text{SA}$. 
First, $J_\text{MFT}$ overestimates the antiferromagnetic contribution [Fig. \ref{Fig:J}(a)]. Thus MFT exchange parameter
is ferromagnetic in a smaller range of $t'/U$ than the exact $J$ and exhibits larger antiferromagnetic contribution than the latter. 
Second, the description of $J_\text{MFT}$ becomes poor when the covalency effects becomes stronger by reducing $U'$ or increasing $t$'s [Figs. \ref{Fig:J}(b)-(d)]. 
Since the magnetic force theorem calculations assume well-localized classical spins, this method would not work well when the covalency effects are strong.

\subsection{Spin-flip approach}

Another single-reference method that allows to maintain the symmetry of the involved spin states is the spin-flip time-dependent DFT approach (SF-TDDFT) \cite{Shao2003}. It starts from a highest-spin configuration within the subsystem of unpaired electrons ($S_{\text{max}}$) and considers a set of one-electron excitations accompanied by the reversal of one electron spin, which after configuration mixing results in a number excited terms with spin $S_{\text{max}}-1$. The closest of them to the reference one can be used for the extraction of exchange parameters. For instance, in the case of two magnetic centers ($A$ and $B$), we have
\begin{equation}
J_{AB} \; S_{\text{max}}= E\big( S_{\text{max}} \big) - E\big( S_{\text{max}} -1 \big) .
\label{Eq:SF}
\end{equation} 
This approach can be applied straightforwardly also to the cases when the state with $S_{\text{max}}$ is not the ground one, which makes it very useful for the calculation of exchange parameters in exchange coupled complexes \cite{Valero2011, Orms2018, Kotaru2023}. It can be equally applied to complexes with a large number of magnetic sites, for which it is especially efficient even in comparison with BS DFT approaches \cite{Mayhall2015}.

\begin{figure}[tb]
\includegraphics[width=0.7\linewidth]{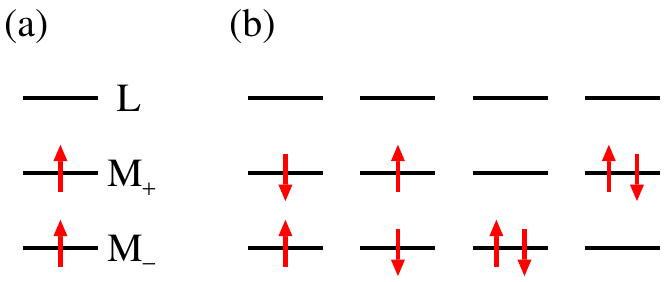}
\caption{
Electron configurations within the spin-flip approach. 
(a) Reference high-spin state. (b) Spin-flip states. 
}
\label{Fig:SF}
\end{figure}

One clear advantage of the SF TDDFT method is the correct description of single-electron excitation energies, which implicitly contain the screening of electronic interaction due to dynamical correlation, first of all, the strong reduction of the $U$ parameter in (\ref{Eq:H}) compared to its bare HF value. Failure to account properly for the screening of $U$ is the major reason for the strong underestimation of antiferromagnetic contribution to $J$ by CASSCF methods. However, the SF TDDFT has an intrinsic drawback of not taking into account many single-electron spin-flip excitations, which may contribute to $J$ Figure \ref{Fig:SF} show the spin-flip excitation from the reference $S=1$ configuration which might be included in a SF TDDFT calculation when applied to the two-site model (\ref{Eq:H}). We see that the requirement of a proper account of spin symmetry of excited singlets rules out completely the excitation involving ligand orbitals [in Fig. \ref{Fig:SF} a hole picture of the ground and SF configurations is employed for simplicity]. This will certainly affect the value of $J$ when the ligand-metal covalency is not small. 

\subsection{Missing contributions to $J$ in the single-reference approaches}

The present analysis shows that the single-reference based approaches fail to provide reliable $J$ values when the magnetic electrons are significantly delocalized over the bridging ligand, a situation realized at sufficiently small $\Delta$. 
Small values of $\Delta$ arise  when the metal and ligand orbitals are close to resonance as, e.g., in iron-sulfur complexes \cite{Noodleman2002, 3center}.
Weaker but non-negligible delocalization occurs in other cases, such as copper-oxygen systems \cite{3center} and iron-oxygen complexes \cite{Postnikov2006}.
Also, as emphasized in Sec. \ref{Sec:model}, finite and even large $t'$ is not surprising in real materials \cite{3center}.

The breakdown of the single-reference approaches for the calculation of $J$ clearly originates from the lack of sufficient flexibility of the trial wave functions.
Indeed, while CI and HF wave functions for ferromagnetic states coincide in our model, they differ for the antiferromagnetic ($S=0$) states through fewer variational parameters in the latter treatment (see Appendix \ref{A:UHF}).
The absence of approximations in the derivation of exact states rules out any doubts that the discrepancy shown by the BS approach is due to its intrinsic drawback.
This consists ultimately in the lack of adequate treatment of electron correlation involving configurations of bridging ligand type.
To eliminate this drawback, the BS approach should be extended as suggested below.

\begin{figure}[tb]
\includegraphics[width=\linewidth]{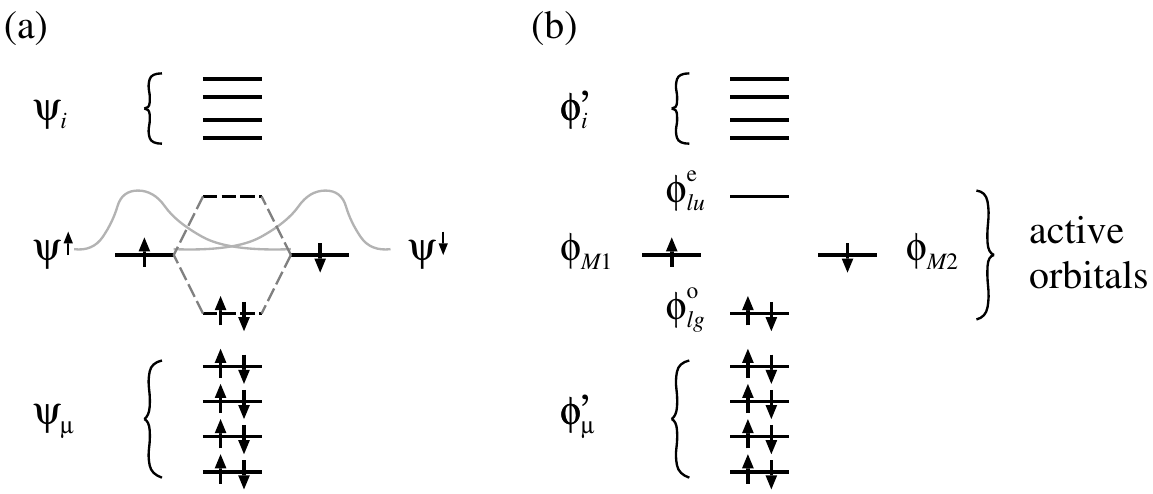}
\caption{
(a) Partially restricted molecular orbitals and (b) active magnetic and ligand orbitals (solid lines). 
}
\label{Fig:config}
\end{figure}

\section{Extension of the BS approach for the exchange parameters}
\label{Sec:MC}
A reliable multiconfigurational calculation of exchange parameters based on Hartree-Fock orbitals, e.g., within the CASSCF/CASPT2 approximation \cite{Malrieu2013}, poses difficulties, especially for large molecules. 
The main drawback of this approach is an insufficient account of dynamical correlation, resulting in an underestimation of the intersite electron 
transfer and an overestimation of on-site electron repulsion. On the contrary, the dynamical correlation is intrinsically contained in DFT, CC and GW, which is the 
reason why the BS approaches became a standard tool for the calculation of $J$. Here we suggest an extension of this approach in order 
to cure its intrinsic drawbacks described above.
To this end, we propose a minimal multiconfigurational version involving electronic configurations built on several active orbitals 
derived from a preliminary BS calculation. To exemplify this methodology, we consider the simplest case of two unpaired electrons in equivalent magnetic (metal) sites, the extension to other situations is straightforward.

\subsection{Spin polarization contribution}
After a standard BS calculation, a partly restricted calculation is performed in which the two magnetic orbitals containing unpaired electrons ($\psi^\sigma$) 
remain unchanged, while the doubly occupied and empty orbitals are reoptimized under spin restriction [Fig. \ref{Fig:config}(a)].
All molecular orbitals are orthogonal to each other except for the magnetic ones:
\begin{eqnarray}
 \langle \psi^\uparrow|\psi^\downarrow \rangle &=& S_{\text{M}}.
\label{Eq:overlap}
\end{eqnarray}
The difference between total energies obtained from BS and partly restricted calculation for the HS and BS states gives the spin polarization contribution 
to the energy of the latter. Accordingly, the spin polarization contribution to the exchange parameter is given by
\begin{eqnarray}
 J_{\text{pol}} &=& J_{\text{BS}}-J_{\text{res}},
\label{Eq:J_pol}
\end{eqnarray}
where $J_{\text{BS}}$ is calculated as in Eq. (\ref{Eq:Jbs}) and $J_{\text{res}}$ is calculated as follows: 
\begin{eqnarray}
 J_{\text{res}} &=& \frac{2(E_{\text{HS}}^{\text{res}} - E_{\text{BS}}^{\text{res}})}{1+S_{\text{M}}^2 }.
\label{Eq:J_rest}
\end{eqnarray}
This equation is a particular case of Yamaguchi's formula (\ref{Eq:Jbs}) when the spin contamination comes from a single pair of non-orthogonal magnetic orbitals \cite{Caballol1997}.
The total exchange parameter consists of the contributions described in the previous section ($J$) and the spin-polarization contribution (\ref{Eq:J_pol}):
\begin{eqnarray}
 J_{\text{tot}} &=& J+J_{\text{pol}}.
\label{Eq:J_tot}
\end{eqnarray}
Note that the inherent error of the BS approach is expected to be almost canceled in $J_{\text{pol}}$ because the energy of the BS configuration enters both 
terms in Eq. (\ref{Eq:J_pol}).

\subsection{The active orbitals}
\label{Sec:active_orbitals}
To overcome the drawbacks of the BS approach, we should be in line with the results of the previous section, i.e. identify a set of effective magnetic and bridging ligand 
orbitals (active orbitals) and undertake a multiconfigurational calculation on their basis. The n from the energies of the lowest $S=0$ and $S=1$ states the parameter $J$ is extracted.
Following the previous section, these active orbitals are chosen in a form that allows the expansion of the BS magnetic orbitals $\psi^\sigma$ 
solely in their basis. The construction of these active orbitals is shown below for the simplest system 
of two equivalent magnetic (metal) sites $i$=1,2 with one unpaired electron, $S_i$=1/2. 
Furthermore, the complex is supposed to possess a mirror symmetry under which the BS orbitals $\psi^\sigma$ pass into each other, 
$\sigma_{\text{h}} 
\psi^\uparrow =\psi^\downarrow$ [Fig. \ref{Fig:config}(a)], whereas the restricted molecular orbitals \{$\psi_{i}, \psi_{\mu}$\} are either even or odd with respect to this 
transformation, i.e., are characterized by indices $g$ and $u$ respectively 
\footnote{In the case when the complex possesses a two-fold rotational symmetry interchanging the magnetic centers, the indices $g$ and $u$ are replaced by the irreducible representations $a_1$ and $a_2$ of the C$_2$ symmetry group, respectively.
}.

For this symmetric complex the orbitals $\psi^\sigma$ are decomposed into four active orbitals (see Appendix \ref{A:MC}):
\begin{eqnarray}
\psi^\uparrow &=& c_1 \phi_{M_1} + c_2 \phi_{M_2} + c^{\text{o}}_g \phi_{lg}^{\text{o}} + c^{\text{e}}_u \phi_{lu}^{\text{e}}, \nonumber\\
\psi^\downarrow &=& c_2 \phi_{M_1} + c_1 \phi_{M_2} + c^{\text{o}}_g \phi_{lg}^{\text{o}} - c^{\text{e}}_u \phi_{lu}^{\text{e}}, 
\label{Eq:psi_sigma}
\end{eqnarray}
where $\phi_{M_1}$ and $\phi_{M_2}$ are magnetic orbitals centered at the metal sites 1 and 2, respectively, accommodating unpaired magnetic electrons 
in the ground electronic configuration [Fig. \ref{Fig:config}(b)]. 
$\phi_{lg}^{\text{o}}$ and $\phi_{lu}^{\text{e}}$ are effective bridging ligand orbitals,
\begin{eqnarray}
\phi_{lg}^{\text{o}} &=& b^{\text{o}}_g \psi_{lg}^{\text{o}} + d^{\text{o}}_g \psi_g, \;\;\;
\phi_{lu}^{\text{e}} = b^{\text{e}}_u \psi_{lu}^{\text{e}} + d^{\text{e}}_u \psi_u,  
\label{Eq:phi_l}
\end{eqnarray}
where $\psi_{lg}^{\text{o}}$ and $\psi_{lu}^{\text{e}}$
are linear combinations of doubly occupied and empty restricted molecular orbitals, respectively [see Fig. \ref{Fig:config}(a)]:
\begin{eqnarray}
\psi_{lg}^{\text{o}} &=& \sum_{\mu} a_{\mu g} \psi_{\mu g} , \;\;\;
\psi_{lu}^{\text{e}} = \sum_{i} a_{i u} \psi_{i u} ,  
\label{Eq:psi_l}
\end{eqnarray}
while $\psi_{g,u}$ are symmetrized combinations of $\psi^\sigma$, 
\begin{eqnarray}
\psi_{g,u} &=& \frac{1}{2(1\pm S_{\text{M}})} \big( \psi^\uparrow \pm \psi^\downarrow \big) .
\label{Eq:psi_o_e}
\end{eqnarray}
One should note that the four orbitals entering the r.h.s. of Eq. (\ref{Eq:phi_l}) are automatically orthogonal, so that the
normality of $\phi_{lg}^{\text{o}}$ and $\phi_{lu}^{\text{e}}$ imposes conventional relations for the expansion coefficients. 
The orthogonality between the orbitals Eqs. (\ref{Eq:psi_sigma}) and (\ref{Eq:phi_l}) gives (see Appendix \ref{A:MC}):
\begin{eqnarray}
c^{\text{o}}_g &=& \frac{d^{\text{o}}_g}{\sqrt{2(1+ S_{\text{M}})}}, \;\;\;  c^{\text{e}}_u= \frac{d^{\text{e}}_u}{\sqrt{2(1- S_{\text{M}})}} ,
\label{Eq:c_o_e}
\end{eqnarray} 
whereas from the normality of $\psi^\sigma$ and orthogonality of $\phi_{M_1}$ and $\phi_{M_2}$ we extract the coefficients $c_1$ and $c_2$, Eq. (\ref{Eq:c_1_2}), 
entering the decomposition (\ref{Eq:psi_sigma}). 
As Eqs. (\ref{Eq:c_o_e}) and (\ref{Eq:c_1_2}) show that all coefficients in this decomposition are expressed via $d^{\text{o}}_g$ and $d^{\text{e}}_u$, which together with the coefficients \{$a_{\mu g}$\} and \{$a_{i u}$\} are the variational parameters defining the two active ligand orbitals in Eq. (\ref{Eq:phi_l}). 
Once these are found from an optimization procedure specified below, all expansion coefficients in Eqs. (\ref{Eq:psi_sigma}) can be calculated, which allows via the knowledge of $\psi^\sigma$ to determine straightforwardly from these equations for the active magnetic orbitals $\phi_{M_1}$ 
and $\phi_{M_2}$ [Eq. (\ref{Eq:M_1_2})]. 

In the case of non-equivalent magnetic centers, the two BS magnetic orbitals $\psi^\sigma$ will not be related by symmetry anymore. Therefore, there is no reason to expect that they will decompose through common active ligand orbitals as in Eq. (\ref{Eq:psi_sigma}) but rather through different ones. Passing from two different active ligand orbitals of each type to their orthogonal combinations, we can decompose $\psi^\uparrow$ and $\psi^\downarrow$ into six common active orbitals: two magnetic, two ligands of doubly occupied type and two ligands of empty type.
This scheme is straightforwardly generalized to several unpaired electrons on magnetic sites and more than two magnetic centers. In the absence of symmetry, we will 
have to define for each magnetic electron ($n$) one magnetic ($\phi_{M_n}$) and two ligands ($\phi_{ln}^{\text{o}}$ and $\phi_{ln}^{\text{e}}$) active 
orbitals. The latter are written in analogy to Eq. (\ref{Eq:phi_l}) as follows:
\begin{eqnarray}
\phi_{ln}^{\text{o}} &=& \sum_{m} \big( b_{nm}^{\text{o}} \psi_{lm}^{\text{o}} + d_{nm}^{\text{o}} \psi_{m} \big) , \nonumber\\
\phi_{ln}^{\text{e}} &=& \sum_{m} \big( b_{nm}^{\text{e}} \psi_{lm}^{\text{e}} + d_{nm}^{\text{e}} \psi_{m} \big) ,  
\label{Eq:phi_l_general} 
\end{eqnarray}
where $\psi_{lm}^{\text{o,e}}$ are suitable combinations of occupied and empty restricted orbitals,
\begin{eqnarray}
\psi_{lm}^{\text{o}} & = & \sum_{\mu} a_{m\mu}^{\text{o}} \psi_{\mu} ,\;\;\; \psi_{lm}^{\text{e}} = \sum_{i} a_{mi}^{\text{e}} \psi_{i}, 
\label{Eq:psi_l_general} 
\end{eqnarray}
and $\psi_{m}$ are arbitrary orthogonal combinations of BS magnetic orbitals \{$\psi_{m}^{\text{BS}}$\}. 
The indices $n$ and $m$ in Eq. (\ref{Eq:phi_l_general}) run over the total number $N_{\text{m}}$ of BS magnetic orbitals 
(usually coinciding with the number of magnetic electrons).  
With knowledge of active ligand orbitals 
(\ref{Eq:phi_l_general}), the active magnetic orbitals $\phi_{M_n}$ are derived from the relations [cf. Eq. (\ref{Eq:psi_sigma})]
\begin{eqnarray}
\psi_{n} &=& \sum_{m} \big( c_{nm} \phi_{M_m} + c_{nm}^{\text{o}} \phi_{lm}^{\text{o}} + c_{nm}^{\text{e}} \phi_{lm} ^{\text{e}} \big) 
\label{Eq:phi_n} 
\end{eqnarray}
for a given set of expansion coefficients \{$c_{nm}, c_{nm}^{\text{o}}, c_{nm}^{\text{e}}$\}.

Besides serving as a tool to calculate $J$, this methodology allows to extract the magnetic and effective ligand orbitals in a strict variational way. 
The existent approaches merely identify them with some Wannier orbitals constructed from arbitrarily chosen group of molecular/band orbitals \cite{wannier_orbitals}, an a priori unjustifiable and often unreliable procedure, especially for large
ligands.

\subsection{A minimal version of MC extension}

Once the active magnetic and bridging ligand orbitals are defined, the lowest spin states are obtained as combinations of the corresponding electronic 
configurations $\Phi_I$: 
\begin{eqnarray}
\Psi &=& \sum_I C_I \Phi_I ,
\label{Eq:Psi_I}
\end{eqnarray} 
where $C_I$ are variational parameters found from a multiconfigurational calculation. 
Another set of variational parameters are the expansion coefficients of the active magnetic and ligand orbitals, Eqs. (\ref{Eq:phi_l_general})-(\ref{Eq:phi_n}).

The main difference from other versions of MC calculations is that now one should apply this treatment not to a group of canonical restricted orbitals, such as $\psi_{\mu }$ and $\psi_{i}$ in Fig. \ref{Fig:config}, but to their linear combinations ($\phi_{ln}^{\text{o}}$, $\phi_{ln}^{\text{e}}$, $\phi_{M_n}$) which are not eigenfunctions of the corresponding mean-field, e.g., Kohn-Sham (KS) operator. 
This implies orthogonalization of all other doubly occupied restricted orbitals to the active ligand orbitals (\ref{Eq:phi_l_general}) or, equivalently, to the orbitals (\ref{Eq:psi_l_general}), i.e. construction of their orthogonal linear combinations \{$\psi'_{\mu }$\} and \{$\psi'_{i}$\}, respectively [Fig. \ref{Fig:config}(b)].  
In practice, the knowledge of the form of the latter (coefficients of their decomposition in terms of original \{$\psi_{\mu }$\} and \{$\psi_{i}$\}) is not needed. 
The reason is the known invariance of the electronic configurations under arbitrary unitary transformations in the space of fully occupied (or fully empty) orbitals \cite{McWeeny_book}, which underlies the following equality of determinants:
\begin{eqnarray}
&&|\psi_{l1}^{\text{o}} \bar{\psi}_{l1}^{\text{o}} \cdots \psi_{lN_{\text{m}}}^{\text{o}} \bar{\psi}_{lN_{\text{m}}}^{\text{o}} 
\psi'_{N_{\text{m}} +1} \bar{\psi'}_{N_{\text{m}} +1} \cdots \psi'_{\mu} 
\bar{\psi'}_{\mu} \cdots \psi'_{N_{\text{d}}} \bar{\psi'}_{N_{\text{d}}} | \nonumber\\
&& = |\psi_{1} \bar{\psi}_{1} \cdots \psi_{\mu} \bar{\psi}_{\mu} \cdots \psi_{N_{\text{d}}} \bar{\psi}_{N_{\text{d}}} | ,
\label{Eq:equivalence} 
\end{eqnarray}
where $N_{\text{d}}$ is the number of doubly occupied restricted orbitals in the DFT calculation.
Since all configurations $\Phi_I$ in Eq. (\ref{Eq:Psi_I}) involve either the l.h.s. determinant as is or with few electrons removed from 
a relatively small number of orbitals $\psi_{ln}^{\text{o}}$, $\bar{\psi}_{ln}^{\text{o}}$, we can describe them via few holes added to the core determinant (\ref{Eq:equivalence}). 
Since, in addition, we are not re-optimizing the orbitals $\psi_{\mu}$ during the MC calculation, the determinant in the r.h.s. represents a ``vacuum function'' for the added holes. 
At the same time, the occupation of BS orbitals (their combinations $\psi_{m}^{\text{BS}}$) and empty restricted orbitals $\psi_{i}$ are described in the electronic representation. 
In this way, the wave functions $\Phi_I$ in the expansion (\ref{Eq:Psi_I}) involve explicitly only a few electrons in the orthogonal BS orbitals and empty restricted orbitals, and a few holes in the doubly occupied restricted orbitals. 
Details of such electron-hole description are given in Appendix \ref{A:MC2}. 

Having established the rules for constructing the electronic configurations $\Phi_I$ in the mixed electron-hole representation, the calculation of the matrix elements $H_{IJ}$ of the corresponding Hamiltonian (\ref{Eq:3-H}) can be done straightforwardly.
Within an explicit version of MC BS calculation
one should minimize the functional:
\begin{eqnarray}
 E = \sum_{I,J} C_I^* C_J H_{IJ}(\{ a_{m\mu}^{\text{o}}, a_{mi}^{\text{e}}, b_{nm}^{\text{o,e}}, d_{nm}^{\text{o,e}},  c_{nm}, c_{nm}^{\text{o,e}} \}), \nonumber\\
\label{Eq:CASSCF} 
\end{eqnarray}
with respect to the CI and orbital coefficients (subject to corresponding orthonormal conditions) in full analogy to a CASSCF calculation \cite{Veryazov_book}.

In this work, we do not discuss the implementation of the proposed approach. One should mention, however, that the feasibility of this scheme depends on the evaluation of off-diagonal matrix elements in Eq. (\ref{Eq:CASSCF}). It is a quite straightforward procedure for HF, CC, and even GW approaches. As for DFT, the evaluation of $H_{IJ}$ can only be done indirectly, within an uncontrolled approximation. Given the popularity of DFT calculations, we review the MC DFT approaches used to date below.

\subsubsection{MC DFT approaches}

Several approaches to the MC DFT have been developed in the past. 
These include a re-definition of Kohn–Sham (KS) theory to include multiconfigurational reference wave function from the start \cite{gorling_2008a,head-gordon_2009a}, a range-separation of the electron-electron interaction into a short-range part described by a local correlation potential and a long-range part described by the correlation arising from the MC expansion \cite{savin_2002a,savin_2012a,jensen_2010a}, 
a method based on a local scaling factor of the DFT correlation energy \cite{cremer_2000d} and more complicated methods for the balanced treatment of MC and DFT correlation effects \cite{cremer_2005a}, 
as well as methods based on a correlation separation using the LDA correlation energy density \cite{yamaguchi_2002c,yamaguchi_2006a,yamaguchi_2007a}. 
As a less rigorous approach, a reparametrization of the XC functional in the context of an MC expansion \cite{truhlar_2004a,truhlar_2005b}, and the rescaling of the matrix elements of the CI matrix constructed in the presence of the XC potential by empirical coefficients \cite{grimme_1999a} have been proposed. 
Another actively developed approach is the so-called ensemble-referenced Kohn-Sham (REKS) method \cite{filatov_1999a,filatov_2000a,filatov_2015a}, in which the variational entity is an ensemble density expanded as a linear combination of densities corresponding to individual determinants.
We note that all these approaches can be applied to our problem in a slightly modified form. 

Concerning Eq. (\ref{Eq:CASSCF}), within DFT, the matrix elements are supposed to be rescaled by empirical coefficients \cite{grimme_1999a}.  
This approach is closely related to the actively used nowadays ROCIS method, a single-configuration multireference approach for evaluation of electronic excitation of inner shells \cite{neese_2013a,neese_2014a}, which also employs the rescaling of CI matrix elements using empirical factors. 
As starting point in the self-consistent calculation, we identify $\phi_{M_n}$ with $\psi_{n}$ which are maximally close to the original $\psi_{m}^{\text{BS}}$, obtained from the latter via e.g. a L\"{o}wdin orthogonalization ($c_{nm}=\delta_{nm}$, $c_{nm}^{\text{o,e}}=0$); 
the orbitals $\phi_{ln}^{\text{o,e}}$ are identified with corresponding restricted orbitals having maximal weight in the pairs of neighbor (overlapping) $\psi_{m}^{\text{BS}}$s. 

Another approach, the so-called CAS-DFT \cite{Cremer2002}, considers the functional  
\begin{eqnarray}
 && E = F^{\text{CAS-DFT}}[\Psi] + E^{\text{CAS-DFT}}_{\text{C}}[\rho ,P] ,
\label{Eq:CAS-DFT} 
\end{eqnarray}
where the first term is the CAS energy (\ref{Eq:CASSCF}) in its conventional form (without rescaling), while the second term is the correlation energy in which the conventional spin densities are replaced by combinations of total CAS density $\rho (\mathbf{r} )$ and on-top pair density $P (\mathbf{r},\mathbf{r} )$ \cite{Becke1995}. 
As appropriate $E^{\text{C}}$, a Colle-Salvetti \cite{CS1975, CS1990} or Lee-Yang-Parr \cite{LYP1988} correlation functional should be used. 
In order to avoid double counting of dynamical correlation energy covered by the first term of (\ref{Eq:CAS-DFT}), $E^{\text{CAS-DFT}}_{\text{C}}[\rho, P]$ is evaluated with local rescaling factors \cite{Cremer2002}. 
This rescaling can be neglected when a few configurations are mixed in the CASSCF wave function, which is certainly the case here. 
Indeed, given the relative weakness of the exchange coupling in most magnetic complexes, we will only need to consider singly and doubly excited configurations from reference one(s), resulting in their limited amount even for a large space of active magnetic and ligand orbitals. 

A related version of CAS-DFT is the actively developed multiconfigurational pair-density functional theory (MC-PDFT) \cite{gagliardi_2014a,gagliardi_2017g}. 
It is currently implemented in OpenMolcas \cite{OpenMolcas@2019} with a plethora of on-top functionals corresponding to translated exchange-correlation functionals and different versions of MC calculations. 
It has been successfully applied to the calculation of relative energy levels in exchange-coupled systems \cite{gagliardi_2019i}, the calculation of singlet-triplet splittings in main-group and organic systems \cite{gagliardi_2016b,gagliardi_2018h,gagliardi_2019b} and the relative spin-state energetics of coordinated metal ions \cite{gagliardi_2019f,gagliardi_2020b}.

Contrary to these methods, which do not involve the KS density, the REKS functional is a linear combination of KS energies corresponding to different electronic configurations of active electrons. 
The coefficients of this combination are expressed via the fractional occupation numbers in the total REKS density through model considerations \cite{filatov_1999a,filatov_2000a,filatov_2015a}. 
The weak point of this approach is that it is designed for tiny active spaces and cannot be easily extended over, e.g., CAS(4,4). 
Note that the latter will be already sufficient for the calculation of exchange parameters in symmetric dimers with one unpaired electron per site, for which the expressions derived in Sec. \ref{Sec:active_orbitals}  and Appendix \ref{A:MC} can be applied directly. 
The densities corresponding to different electronic configurations of active electrons are calculated as described in Appendix \ref{A:MC2}. 
As an example, Eq. (\ref{Eq:Phi_1-density}) gives the total density for the ground configuration $\Phi_1$. 
One should keep in mind that the self-consistent procedure involves a variation of orbital coefficients defining the active orbitals only. 
The same refers also to other approaches mentioned above.

One should note that the MC DFT approaches have been straightforwardly applied to the calculation of exchange parameters in organic materials \cite{yamaguchi_2002b,yamaguchi_2007a} and complexes (an overview of earlier work can be found in Ref. \onlinecite{illas_2007b}). 
They generally produced results comparable in accuracy with BS DFT calculations. 
Thus, CASSCF(2,2) calculations of magnetic coupling in Cu(II) binuclear complexes by the REKS method have shown that imposing strict spin symmetry does not improve the BS DFT evaluation of exchange parameters \cite{illas_2007a}. 
While the active space in these calculations was restricted to magnetic orbitals only, we stress that including specially designed active ligand orbitals is expected to improve the predictability of $J$ in such calculations, as they would certainly improve the BS DFT results according to the present study.

\section{Conclusion}
In this work, we prove the breakdown of the broken-symmetry approach for the evaluation of exchange parameters by applying it to a generic three-site model. 
We show that this breakdown originates from an artificial constraint on the multiconfigurational state imposed by the broken-symmetry determinant. It is also found that other single-reference based approaches do not cure this drawback.
The error becomes significantly pronounced in the case of strong covalency between magnetic centers and the bridging ligand.
To cure this drawback, we propose a calculational scheme based on a minimal multiconfigurational extension of the BS approach. 
An example of such an economical employment of the CI space for the description of realistic systems is the recently developed GS-ROCIS method \cite{Leyser2024}. 

As active orbitals in the MC calculations, the proposed method employs effective magnetic and bridging ligand-type orbitals, whose construction and self-consistent determination are outlined in detail. 
This approach can be used with a variety of quantum chemistry software involving MC and BS calculations, in particular, with any version of the existing MC DFT code, the only required modification being the implementation of optimization of the coefficients defining the active orbitals. 
Besides possible improvement in the prediction of exchange parameters, we expect this approach to help resolve the issue related to the strong variation of the performance of a given exchange-correlation functional for evaluation of $J$ in different magnetic systems \cite{Ruiz1997, Akseli_PhD}, which variability seems to be less pronounced for other molecular properties calculated with DFT.

\section*{Acknowledgement}
N.I. was supported by Grant-in-Aid for Scientific Research (Grant No. 22K03507) from the Japan Society for the Promotion of Science, and Chiba University Open Recruitment for International Exchange Program.
Z.H. was supported by the China Scholarship Council, and the financial support of research projects A-8000709-00-00, A-8000017-00-00, and A-8001894-00-00 of the National University of Singapore. 
A.M. acknowledges funding provided by the Research Council of Finland (grant no. 362649).

\section*{Author Declarations}
\subsection*{Conflict of Interest}
The authors have no conflicts to disclose.

\subsection*{Author Contributions}
N. Iwahara: Conceptualization (equal); Formal analysis (lead); Investigation (equal); Writing – original draft (equal); Writing – review \& editing (equal).
Z. Huang: Writing – review \& editing (equal).
A. Mansikkam\"{a}ki: Writing – review \& editing (equal).
L. F. Chibotaru: Conceptualization (equal); Project administration (lead); Investigation (equal); Writing – original draft (equal); Writing – review \& editing (equal). 

\section*{Data Availability}
The data that support the findings of this study are available from the corresponding authors upon request.

\appendix

\section{Solutions for the generic three-site model}
The full CI and HF treatments of the three-center Hubbard model (\ref{Eq:H}) are shown below. 

\subsection{Exact solutions}
\label{A:exact}

\subsubsection{Ferromagnetic state}
The basis for the ferromagnetic (spin-triplet, $S = 1$) states $|\text{F}, M_S; n, p\rangle$ are
\begin{eqnarray}
 |\text{F},1;0-\rangle &=& |12\rangle, 
\quad
 |\text{F},1;1\mp \rangle = \frac{1}{\sqrt{2}} (|1l\rangle \pm |l2\rangle),
\label{Eq:F1}
\end{eqnarray}
and 
\begin{eqnarray}
 |\text{F},0;0-\rangle &=& \frac{1}{\sqrt{2}}(|1\bar{2}\rangle - |2\bar{1}\rangle),
\nonumber\\
 |\text{F},0;1\mp\rangle &=& \frac{1}{2}(|1\bar{l}\rangle - |l\bar{1}\rangle \pm |l\bar{2}\rangle \mp |2\bar{l}\rangle),
\label{Eq:F0}
\end{eqnarray}
respectively.
Here, $|ij\rangle$ and $|i\bar{j}\rangle$ etc. indicate Slater determinants, spin up and down are specified by without and with bar,
``F'' stands for ferromagnetic state, $M_S$ the $z$ component of the total spin, 
$n$ distinguishes the states characterized by the same $S$ and $M_S$, $p$ ($= \pm $) the parity of the spatial part (symmetric or antisymmetric).

The Hamiltonian matrix is written as 
\begin{eqnarray}
 \mathbf{H}_\text{F} &=& 
 \begin{pmatrix}
  0 & \sqrt{2}t & 0 \\
  \sqrt{2}t & \Delta - t' & 0 \\
  0 & 0 & \Delta + t'
 \end{pmatrix},
 \label{Eq:HmatF}
\end{eqnarray}
with the ferromagnetic basis in the order of $|0-\rangle$, $|1-\rangle$, $|1+\rangle$ (``F'' and $M_S$ are omitted).
The ground energy is obtained from the $2 \times 2$ block of Eq. (\ref{Eq:HmatF}), and the ground state is expressed as
\begin{eqnarray}
 |\Psi^\text{F}_{M_S} \rangle &=& \sum_{i=0,1} |\text{F},M_S;i-\rangle C_i.
\label{Eq:Psi_F}
\end{eqnarray}

\subsubsection{Antiferromagnetic state}
The symmetrized antiferromagnetic (spin-singlet, $S=0$) states $|\text{AF}; n, p\rangle$ are 
\begin{eqnarray}
 |\text{AF};0+\rangle &=& \frac{1}{\sqrt{2}} (|1\bar{2}\rangle + |2\bar{1}\rangle),
\nonumber\\
 |\text{AF};1\pm \rangle &=& \frac{1}{2} (|1\bar{l}\rangle + |l\bar{1}\rangle \pm |l\bar{2}\rangle \pm |2\bar{l}\rangle),
\nonumber\\
 |\text{AF};2\pm \rangle &=& \frac{1}{\sqrt{2}} (|1\bar{1}\rangle \pm |2\bar{2}\rangle),
\nonumber\\
 |\text{AF};3+ \rangle &=& |l\bar{l}\rangle.
\label{Eq:AF}
\end{eqnarray}
The Hamiltonian matrix is written as 
\begin{eqnarray}
 \mathbf{H}_\text{AF} &=& 
 \begin{pmatrix}
  0 & \sqrt{2}t & 2t' & 0 & 0 & 0 \\
  \sqrt{2}t & \Delta + t' & \sqrt{2}t & 2t & 0 & 0 \\
  2t' & \sqrt{2}t & U & 0 & 0 & 0 \\
  0 & 2t & 0 & 2\Delta + U'& 0 & 0 \\
  0 & 0 & 0 & 0 & \Delta - t' & \sqrt{2}t \\
  0 & 0 & 0 & 0 & \sqrt{2}t & U 
 \end{pmatrix},
\nonumber\\
\label{Eq:HAF}
\end{eqnarray}
in the order of the basis $|0+\rangle$, $|1+\rangle$, $|2+\rangle$, $|3+\rangle$, $|1-\rangle$, $|2-\rangle$
(``AF'' is omitted). 
The ground energy is obtained from the spatially symmetric part (the $4\times 4$ block). The ground antiferromagnetic state is written as 
\begin{eqnarray}
 |\Psi^\text{AF} \rangle &=& \sum_{i=0}^3 |\text{AF};i+\rangle C_i.
\label{Eq:Psi_AF}
\end{eqnarray}
The exact antiferromagnetic ground state of the present model (\ref{Eq:H}) has three parameters considering the normalization condition.

\subsection{Hartree-Fock solutions}
\label{A:HF}
\subsubsection{Ferromagnetic state}
\label{A:ferro}
From the atomic orbitals, $|1\rangle$, $|2\rangle$, and $|l\rangle$, two symmetric (S) and one antisymmetric (A) molecular orbitals are constructed: 
\begin{eqnarray}
 |\psi_\text{S}\rangle &=& \frac{A}{\sqrt{2}}(|1\rangle + |2\rangle) + B |l\rangle,
\nonumber\\
 |\psi'_\text{S}\rangle &=& \frac{B}{\sqrt{2}}(|1\rangle + |2\rangle) - A |l\rangle,
\nonumber\\
 |\psi_\text{A}\rangle &=& \frac{1}{\sqrt{2}}(|1\rangle - |2\rangle).
\label{Eq:MO}
\end{eqnarray}
The coefficients $A$ and $B$ are real, and the molecular orbitals are normalized. 
The high-spin state with maximal projection $M_S = 1$,
\begin{eqnarray}
 |\Psi_\text{HF}^\text{F}\rangle &=& |\psi_\text{S}\psi_\text{A}\rangle
= -A |\text{F},1;0-\rangle -B |\text{F},1;1-\rangle.
\label{Eq:Slater_Ferro}
\end{eqnarray}
Since both the high-spin full CI and HF wave functions contain one variational parameter, the HF wave function can describe the exact high-spin state.

\subsubsection{Broken symmetry low-spin state}
\label{A:UHF}
The molecular orbitals used for the BS-HF (or unrestricted HF, UHF) method are written as 
\begin{eqnarray}
 |\psi^\uparrow\rangle &=& C_1|1\rangle + C_2|2\rangle + C_l|l\rangle,
\nonumber\\
 |\psi^\downarrow\rangle &=& C_2|1\rangle + C_1|2\rangle + C_l|l\rangle,
\label{Eq:MO_UHF}
\end{eqnarray}
where the coefficients $C_1, C_2, C_l$ are real, and the molecular orbitals are normalized.
The BS-HF wave function is
\begin{eqnarray}
 |\Psi_\text{BS}\rangle &=& |\psi^\uparrow \bar{\psi}^\downarrow \rangle
 \label{Eq:Eq:WF_UHF}
 \\
 &=& \frac{C_1^2+C_2^2}{\sqrt{2}}|\text{AF};0+\rangle 
 + (C_1+C_2)C_l|\text{AF};1+\rangle 
\nonumber\\
 &&+ \sqrt{2}C_1C_2|\text{AF};2+\rangle
 + C_l^2|\text{AF};3+\rangle
\nonumber\\
 &&+ \frac{C_1^2-C_2^2}{\sqrt{2}}|\text{F},0;0-\rangle
 + (C_1-C_2)C_l|\text{F},0;1-\rangle.
\nonumber\\
\label{Eq:WF_UHF_config}
\end{eqnarray}
This expression shows that both ferro- and antiferromagnetic configurations are included in $|\Psi_\text{BS}\rangle$. 
Note that the exact low-spin states have three variational parameters, while the BS states contain only two parameters. 

Based on the wave function (\ref{Eq:WF_UHF_config}), the BS-HF energy and $\langle \hat{\bm{S}}^2 \rangle_\text{BS}$ are calculated as, respectively,
\begin{eqnarray}
 E_\text{BS} &=& 4t C_l(C_1+C_2) + 4t' C_1 C_2  + 2\Delta C_l^2 
\nonumber\\
 &&+ 2U C_1^2C_2^2 + U' C_l^4,
\label{Eq:EUHF}
\\
 \langle \hat{\bm{S}}^2 \rangle_\text{BS} &=& 
 \langle \hat{\bm{S}}^2 \rangle_\text{F} 
  \left[
 \frac{(C_1^2-C_2^2)^2}{2}
 + (C_1-C_2)^2C_l^2
 \right].
\qquad
\label{Eq:S2UHF_2}
\end{eqnarray}
We determine the molecular orbital coefficients by numerically minimizing $E_\text{BS}$.
The calculated $\langle \hat{\bm{S}}^2 \rangle_\text{BS}$ is shown in Fig. \ref{Fig:S2}.

\subsubsection{Spin symmetry adapted state}
\label{A:SA}
The spin symmetry adapted (SA) HF wave function is the AF ($S=0$) part of the BS wave function (\ref{Eq:WF_UHF_config}):
\begin{eqnarray}
 |\Psi_\text{SA}\rangle &=& \frac{1}{\sqrt{2(1+S_M^2)}} \left(|\psi^\uparrow \bar{\psi}^\downarrow \rangle + |\psi^\downarrow \bar{\psi}^\uparrow \rangle\right)
\label{Eq:WF_SA}
\\
 &=&
 \frac{1}{\sqrt{1+S_M^2}} 
 \Big[
 (C_1^2+C_2^2)|\text{AF};0+\rangle 
\nonumber\\
 &+&
   \sqrt{2}(C_1+C_2)C_l |\text{AF};1+\rangle
 + 2C_1C_2|\text{AF};2+\rangle 
\nonumber\\
 &+&
   \sqrt{2}C_l^2|\text{AF};3+\rangle 
 \Big],
\label{Eq:WF_SA_config}
\end{eqnarray}
where $S_M = \langle \psi^\uparrow|\psi^\downarrow \rangle = 2C_1C_2+C_l^2$. 
Since the right-hand side contains only the AF configurations, the spin expectation value for $|\Psi_\text{SA}\rangle$ is 0. 

The energy for the SA-HF state is
\begin{eqnarray}
 E_\text{SA} &=& \frac{1}{1+S_M^2} 
 \Big[
  4t(1+S_M)(C_1+C_2) C_l 
\nonumber\\
 && 
 + t'(4C_1C_2+2(C_1^2+C_2^2)S_M) 
\nonumber\\
 &&
 + 2 \Delta C_l^2(1+S_M) 
 + 4UC_1^2C_2^2 
 + 2U'C_l^4
 \Big].
\label{Eq:E_SA}
\end{eqnarray}
We obtain the molecular orbital coefficients by minimizing $E_\text{SA}$.

\section{Expressions for the magnetic force theorem}
\label{A:MFT}
Here, we show the detailed expressions used for the magnetic force theorem calculations of $J_\text{MFT}$. 
The rotations of the quantum spins of the broken symmetry state (\ref{Eq:Eq:WF_UHF}) results in 
\begin{align}
 \hat{R}(\theta) |\psi^\uparrow \bar{\psi}^\downarrow\rangle
 &= 
 \cos^2\frac{\theta}{2} |\psi^\uparrow \bar{\psi}^\downarrow\rangle - \sin^2\frac{\theta}{2} |\psi^\downarrow \bar{\psi}^\uparrow\rangle
 \nonumber\\
 &-\frac{1}{2} \sin \theta \left(|\psi^\uparrow \psi^\downarrow\rangle + |\bar{\psi}^\uparrow \bar{\psi}^\downarrow\rangle\right).
\end{align}
Here, $\hat{R}(\theta)$ corresponds to Eq. (\ref{Eq:Rtheta}).
In terms of the configurations, the rotated broken symmetry state is 
\begin{align}
 \hat{R}(\theta) |\psi^\uparrow \bar{\psi}^\downarrow\rangle
 &= 
 \cos\theta |\psi^\uparrow \bar{\psi}^\downarrow\rangle
 - \sum_{M_S=\mp 1} \frac{1}{2}\sin \theta 
 \left[
  \left(C_1^2-C_2^2\right) 
  \right.
 \nonumber\\
 &\times
 \left.
  |\text{F},M_S;0\rangle + \sqrt{2} (C_1-C_2)C_l |\text{F},M_S;1-\rangle
 \right].
\end{align}
The energy expectation value for $\hat{R}(\theta) |\psi^\uparrow \bar{\psi}^\downarrow\rangle$ is, for small $\theta$, 
\begin{align}
 E(\theta) &\approx E_\text{BS} + \theta^2  
 \left[
  -4(C_1+C_2)C_l(2C_1C_2 + C_l^2) t 
  \right.
\nonumber\\
  &- 
  2(C_1^2+C_2^2)(2C_1C_2 + C_l^2) t'
\nonumber\\
  &- 
  \left.
  2(C_1C_2+C_l^2)C_l^2 \Delta -2C_1^2C_2^2 U - C_l^4 U'
 \right].
 \label{Eq:Eq_theta}
\end{align}
Comparing Eqs. (\ref{Eq:Ecl_theta}) and (\ref{Eq:Eq_theta}), we obtain 
\begin{align}
 J_\text{MFT} &= 2
 \left[
  -4(C_1+C_2)C_l(2C_1C_2 + C_l^2) t 
  \right.
\nonumber\\
  &- 
  2(C_1^2+C_2^2)(2C_1C_2 + C_l^2) t'
\nonumber\\
  &- 
  \left.
  2(C_1C_2+C_l^2)C_l^2 \Delta -2C_1^2C_2^2 U - C_l^4 U'
 \right].
 \label{Eq:JMFT}
\end{align}

\section{Details of minimal MC calculation}
\label{A:MC}
\subsection{Decomposition of $\psi^\sigma$ into active orbitals for the generic thee-site model}
\label{A:MC1}
The general decomposition of $\psi^\uparrow$ and $\psi^\downarrow$ should look as follows:
\begin{eqnarray}
\psi^\uparrow &=& c_1 \phi_{M_1} + c_2 \phi_{M_2} + c^{\text{o}}_g \phi_{lg}^{\text{o}} + c^{\text{o}}_u \phi_{lu}^{\text{o}} + 
c^{\text{e}}_g \phi_{lg}^{\text{e}} + c^{\text{e}}_u \phi_{lu}^{\text{e}}, \nonumber\\
\psi^\downarrow &=& c_2 \phi_{M_1} + c_1 \phi_{M_2} + c^{\text{o}}_g \phi_{lg}^{\text{o}} - c^{\text{o}}_u \phi_{lu}^{\text{o}} 
+ c^{\text{e}}g \phi_{lg}^{\text{e}} - c^{\text{e}}_u \phi_{lu}^{\text{e}}, \nonumber\\
\label{Eq:psi_sigma_complete}
\end{eqnarray}
where, compared to Eq. (\ref{Eq:psi_sigma}), the ungerade doubly occupied and gerade empty ligand active orbitals have been added for completeness:
\begin{eqnarray}
\phi_{lu}^{\text{o}} &=& \sum_{\mu} a_{\mu u} \psi_{\mu u} , \;\;\;
\phi_{lg}^{\text{e}} = \sum_{\mu} a_{i g} \psi_{i g} .  
\label{Eq:phi_l_add}
\end{eqnarray}
Taking into account the orthonormality of the active ligand orbitals entering Eq. (\ref{Eq:psi_sigma_complete}) and their orthogonality to $\phi_{M_1}$ 
and $\phi_{M_2}$, we calculate their overlaps with $\psi^\sigma$ from which the expansion coefficients $c^{\text{o}}_u$ and $c^{\text{e}}_g$ are found to 
be zero, while $c^{\text{o}}_g$ and $c^{\text{e}}_u$ are given by Eq. (\ref{Eq:c_o_e}).

The expansion coefficients $c_1$ and $c_2$ in Eq. (\ref{Eq:psi_sigma_complete}) are found by imposing the orthonormality on $\phi_{M_1}$ and $\phi_{M_2}$:
\begin{eqnarray}
c_{1,2} &=& \frac{1}{2} \bigg( \sqrt{B + A} \pm \sqrt{B - A} \bigg) , \nonumber\\
B &=& 1 - {c^{\text{o}}_g}^2 - {c^{\text{e}}_u}^2 , \nonumber\\
A &=& \frac{S_{\text{M}} - {c^{\text{o}}_g}^2 + {c^{\text{e}}_u}^2}{B} .
\label{Eq:c_1_2}
\end{eqnarray}
Thus, having in mind the relations (\ref{Eq:c_o_e}), all expansion coefficients in Eq. (\ref{Eq:psi_sigma_complete}) depend on $d^{\text{o}}_g$ and $d^{\text{e}}_u$.

Using Eq. (\ref{Eq:c_1_2}) we derive the active magnetic orbitals:
\begin{eqnarray}
 \phi_{M_{1,2}} &=& 
 \frac{\big( 1+ S_{\text{M}} - {d^{\text{o}}_g}^2 \big) \; \psi_{\text{o}} -b^{\text{o}}_g d^{\text{o}}_g \; \psi_{lg}^{\text{o}}}{(c_1 + c_2 ) \sqrt{2 ( 1+S_{\text{M}} )}} 
\nonumber\\
&&\pm \frac{\big( 1- S_{\text{M}} - {d^{\text{e}}_u}^2 \big) \; \psi_{\text{e}} -d^{\text{e}}_u b^{\text{e}}_u \; \psi_{lu}^{\text{e}}}{(c_1 - c_2 ) \sqrt{2 ( 1-S_{\text{M}} )}} .
\label{Eq:M_1_2}
\end{eqnarray} 
Given the relations $b^{\text{o}}_g =\sqrt{1-{d^{\text{o}}_g}^2}$, $b^{\text{e}}_u =\sqrt{1-{d^{\text{e}}_u}^2}$ [see Eq. (\ref{Eq:phi_l})] and Eqs. (\ref{Eq:c_o_e}), the orbitals $\phi_{M_{1,2}}$ are 
defined only through the coefficients $d^{\text{o}}_g$ and $d^{\text{e}}_u$.

\subsection{MC calculation in the electron-hole representation}
\label{A:MC2}
For the restricted doubly occupied orbitals $\psi_{\mu}$, we pass from the electron to the hole representation. In the language of second quantization
\cite{McWeeny_book} the electron creation is replaced by hole annihilation and vice versa:
\begin{eqnarray}
&& b_{\mu\sigma} = a_{\mu\sigma}^{\dagger} , \;\;\; b_{\mu\sigma}^{\dagger}  = a_{\mu\sigma} .
\label{Eq:second_quantization}
\end{eqnarray} 
\subsubsection{Configuration functions in the electron-hole representation}
Considering the determinant of restricted doubly occupied states, Eq. (\ref{Eq:equivalence}), as a vacuum function with respect to added holes (removed electrons) 
, $|0\rangle_h$, while keeping the usual electronic representation for magnetic and restricted empty orbitals, with the vacuum function 
with respect to added electrons to these orbitals, $|0\rangle_e$, we can represent any determinant entering the configuration functions $\Phi_I$ in  
(\ref{Eq:Psi_I}) as products of few $a_{i\sigma}^{\dagger}$ and $b_{\mu\sigma}^{\dagger}$ operators acting on the vacuum function $|0\rangle_e |0\rangle_h$
 ($\equiv |0\rangle$). The total number of these operators can differ from one product to another, however, the difference of electronic and hole creation 
operators is equal to $N_{\text{m}}$ for each product.

For instance, the ground configuration in Fig. \ref{Fig:config}(b), described by the determinant
\begin{eqnarray}
&& \Phi_1 = |\phi_{M_1} \bar{\phi}_{M_2} \phi_{lg}^{\text{o}} \bar{\phi}_{lg}^{\text{o}}  
\psi'_{1} \bar{\psi'}_{1} \cdots \psi'_{\mu} 
\bar{\psi'}_{\mu} \cdots | ,
\label{Eq:Phi_1-determinant} 
\end{eqnarray}
is written in the electron-hole representation as follows
\begin{eqnarray}
\Phi_1 =  && \big( \alpha a^{\dagger}_{\text{o}\uparrow} +\beta b_{l\text{o}\uparrow} +\gamma a^{\dagger}_{\text{e}\uparrow} +
\delta a^{\dagger}_{l\text{e}\uparrow} \big) \nonumber\\
&& \big( \alpha a^{\dagger}_{\text{o}\downarrow} +\beta b_{l\text{o}\downarrow} -\gamma a^{\dagger}_{\text{e}\downarrow} -
\delta a^{\dagger}_{l\text{e}\downarrow} \big) 
\big( b^{\text{o}}_g \beta b_{l\text{o}\uparrow} + d^{\text{o}}_g a^{\dagger}_{\text{o}\uparrow} \big) \nonumber\\
&& \big( b^{\text{o}}_g \beta b_{l\text{o}\downarrow} + d^{\text{o}}_g a^{\dagger}_{\text{o}\downarrow} \big)  
b^{\dagger}_{l\text{o}\uparrow} b^{\dagger}_{l\text{o}\downarrow} |0\rangle,
\label{Eq:Phi_1-second_quantization} 
\end{eqnarray}
where $\alpha, \beta, \gamma, \delta$ are the coefficients in front of the corresponding orbital functions in (\ref{Eq:M_1_2}), the latter being replaced 
by corresponding electron creation and hole annihilation operators (the parity index was dropped for shortness); the coefficients $d^{\text{o}}_g$ and $b^{\text{o}}_g$ are the 
same as in Eq. (\ref{Eq:phi_l}).

The total density corresponding to $\Phi_1$ can be written after making use of the relation (\ref{Eq:equivalence}) in the following form: 
\begin{eqnarray}
\rho^{\Phi_1} (\mathbf{r}) = &&\sum_{\mu}^{\text{d.occ}} 2|\psi_{\mu} (\mathbf{r}) |^2 - 2|\psi_{lg}^{\text{o}}(\mathbf{r})|^2 
+ 2|\phi_{lg}^{\text{o}}(\mathbf{r})|^2 \nonumber\\
&&+|\phi_{M_1} (\mathbf{r})|^2 + |\phi_{M_2} (\mathbf{r})|^2 ,
\label{Eq:Phi_1-density}
\end{eqnarray}
where the expressions of active orbitals via restricted and BS molecular orbitals are given by Eqs. (\ref{Eq:phi_l})-(\ref{Eq:psi_o_e}) and 
(\ref{Eq:M_1_2}). 

\subsubsection{The Hamiltonian in the electron-hole representation}
In the electronic Hamiltonian 
\begin{eqnarray}
&& \hat{H} = \sum_{ij\sigma} h_{ij} a^{\dagger}_{i\sigma} a_{j\sigma} +
\frac{1}{2} \sum_{ijkl} \sum_{\sigma\sigma'} V_{ijkl} a^{\dagger}_{i\sigma} a^{\dagger}_{j\sigma'} a_{l\sigma'} a_{k\sigma} 
\label{Eq:H_a}
\end{eqnarray}
we pass to hole operators (\ref{Eq:second_quantization}) for restricted doubly occupied orbitals $\psi_{\mu}$ and obtain:
\begin{eqnarray}
&& \hat{H} = \hat{H}_a + \hat{H}_b + \hat{H}_{ab} + E_{\text{d.occ}} ,
\label{Eq:3-H}
\end{eqnarray}
where $\hat{H}_a$ is the electronic part given by Eq. (\ref{Eq:H_a}) in which the summations over orbital indices exclude the orbitals $\psi_{\mu}$, and
$\hat{H}_b$ is the Hamiltonian for holes (in the following formulas, the hole orbitals are denoted by Greek letters):
\begin{eqnarray}
\hat{H}_b =&&  -\sum_{\mu\nu\sigma} \big[ h_{\nu\mu} + \sum_{\kappa} \big( 2 V_{\nu\kappa\mu\kappa} - V_{\nu\kappa\kappa\mu} \big) \big] 
b^{\dagger}_{\mu\sigma} b_{\nu\sigma} \nonumber\\
&& +\frac{1}{2} \sum_{\mu\nu\kappa\rho} \sum_{\sigma\sigma'} V_{\rho\kappa\nu\mu} b^{\dagger}_{\mu\sigma} b^{\dagger}_{\nu\sigma'} 
b_{\rho\sigma'} b_{\kappa\sigma} .
\label{Eq:H_b}
\end{eqnarray}
The third term in (\ref{Eq:3-H}) is the mixed electron-hole part,
\begin{eqnarray}
&& \hat{H}_{ab} = \sum_{i\mu\sigma} h_{i\mu} \big( b_{\mu\sigma} a_{i\sigma} + a^{\dagger}_{i\sigma} b^{\dagger}_{\mu\sigma} \big) \nonumber\\
&& + \sum_{ijk\mu} \sum_{\sigma\sigma'} V_{ij\mu k} \big( a^{\dagger}_{i\sigma} a^{\dagger}_{j\sigma'} a_{k\sigma'} b^{\dagger}_{\mu\sigma} 
+ b_{\mu\sigma} a^{\dagger}_{k\sigma'} a_{j\sigma'} a_{i\sigma} \big) \nonumber\\
&& + \sum_{i\mu\nu\kappa}\sum_{\sigma\sigma'} V_{ij\mu\kappa} \big( a^{\dagger}_{i\sigma} b_{\mu\sigma'} b^{\dagger}_{\nu\sigma'} b^{\dagger}_{\kappa\sigma} 
+ b_{\kappa\sigma} b_{\nu\sigma'} b^{\dagger}_{\mu\sigma'} a_{i\sigma} \big) \nonumber\\
&& + \frac{1}{2} \sum_{ij\mu\nu}\sum_{\sigma\sigma'} V_{ij\mu\nu} \big( a^{\dagger}_{i\sigma} a^{\dagger}_{j\sigma'} b^{\dagger}_{\nu\sigma'} b^{\dagger}_{\mu\sigma}  
+ b_{\mu\sigma} b_{\nu\sigma'} a_{j\sigma'} a_{i\sigma} \big) \nonumber\\
&& + \sum_{ij\mu\nu}\sum_{\sigma\sigma'} \big( V_{\mu i\nu j}  a^{\dagger}_{i\sigma} a_{j\sigma} b_{\mu\sigma'} b^{\dagger}_{\nu\sigma'} 
- V_{i\mu\nu j} a^{\dagger}_{i\sigma} a_{j\sigma'} b_{\mu\sigma'} b^{\dagger}_{\nu\sigma} \big) , \nonumber\\
\label{Eq:H_ab}
\end{eqnarray}
and the last one is the energy of the closed shell of restricted doubly occupied orbitals:  
\begin{eqnarray}
E_{\text{d.occ}} =&&  2\sum_{\mu} h_{\mu\mu} + \sum_{\mu\mu'} \big( 2 V_{\mu\mu'\mu\mu'} - V_{\mu\mu'\mu'\mu} \big) . \;\;
\label{Eq:E_d.occ}
\end{eqnarray}
The above expressions were derived for real orbitals, implying usual symmetry relations for matrix elements (indices refer to all orbitals:
\begin{eqnarray}
h_{ij} = h_{ji}, \;\; V_{ijkl}=V_{kjil}= V_{ilkj}=V_{klij}=\cdots . \;\;
\label{Eq:symmetry_m.e.}
\end{eqnarray}
In terms of canonical KS orbitals, the electronic Hamiltonian is obtained from (\ref{Eq:H_a}) via the following replacements:
\begin{eqnarray}
i) \;\; && h_{ij} \rightarrow \epsilon_i \delta_{ij}, \nonumber\\ 
ii) \;\; &&\frac{1}{2} \sum_{ijkl} \sum_{\sigma\sigma'} V_{ijkl} a^{\dagger}_{i\sigma} a^{\dagger}_{j\sigma'} a_{l\sigma'} a_{k\sigma} \rightarrow \nonumber\\ 
&&\frac{1}{2} \sum_{ijkl} \sum_{\sigma\sigma'} V_{ijkl} \big( 1-2\delta_{ij}\big) a^{\dagger}_{i\sigma} a^{\dagger}_{j\sigma'} a_{l\sigma'} a_{k\sigma}
\nonumber\\
&& -\sum_{ij\sigma} \big[ \big( v_{\text{c}} \big)_{ij} - \big( v_{\text{x}} \big)_{ij} \big] a^{\dagger}_{i\sigma} a_{j\sigma} ,
\label{Eq:KS}
\end{eqnarray}
where is the KS orbital energy and $v_{\text{c}}$ and $v_{\text{x}}$ is the correlation and exchange potential, respectively. The electron-hole 
representation for this Hamiltonian is derived similarly to Eqs. (\ref{Eq:3-H})-(\ref{Eq:E_d.occ}).

%\bibliography{ref} 

%aipnum4-2.bst 2019-01-14 (MD) hand-edited version of apsrev4-1.bst

\end{document}